# Bioinformatic Scaling of Allosteric Interactions in Biomedical Isozymes

J. C. Phillips
Dept. of Physics and Astronomy, Rutgers University, Piscataway, N. J., 08854

**Abstract**

Allosteric (long-range) interactions can be surprisingly strong in proteins of biomedical interest. Here we use bioinformatic scaling to connect prior results on nonsteroidal anti-inflammatory drugs to promising new drugs that inhibit cancer cell metabolism. Many parallel features are apparent, which explain how even one amino acid mutation, remote from active sites, can alter medical results. The enzyme twins involved are cyclooxygenase (aspirin) and isocitrate dehydrogenase (IDH). The IDH results are accurate to 1% and are overdetermined by adjusting a single bioinformatic scaling parameter. It appears that the final stage in optimizing protein functionality may involve leveling of the hydrophobic cutoffs of the arms of conformational hydrophilic hinges.

## Introduction

Evolution has produced exponentially complex proteins that are not easily described by simple algebraic models suitable for Newtonian molecular dynamic simulations. The value of Newtonian models is especially limited in the context of protein complex formation (dimers, dimers of dimers (tetramers), …). Mutated amino acid interactions at a distance (allosteric interactions) in such complexes are a special challenge. Myoglobin (Mb) was the first protein to have its three-dimensional structure revealed by X-ray crystallography [1], and Mb has been described as an allosteric enzyme [2]. A search on the Web of Science for famous papers (>200 citations) on enzymes and allostery yielded seven articles, including one attempting to explain distant interactions through lowest order Newtonian nonlinear elasticity [3]. This led to the conclusion that nonlinear mechanisms fail, as all strain interactions are local and short-range, unless there is a crack - an undefined, exponentially unlikely modular conformational change.

Many large-scale conformational protein motions are associated with hinges [4,5]. Some of these hinges are located at or near turns in static Euclidean structures determined crystallographically, but this correspondence is not unique, as some turns are not hinges, and vice-versa. Conformational motions are often associated with low-frequency ("soft") elastic modes [6], but the calculation of such modes can depend on details of the force-field used, as well as the hydrogen bonds to a monolayer water film covering the protein [7]. While there is already a large literature on this subject, the bioinformatic scaling method used here is novel, and it appears to be very accurate – accurate enough for biomedical applications.

Hinges appear naturally in both elastic and hydropathic profile models, in the latter as deep hydrophilic minima. The hydropathic profiles contain two kinds of extrema, hydrophilic minima and hydrophobic maxima. Our limited experience so far suggests that hydrophobic maxima are often biomedically critical, and their effects are less evident in elastic models. Each elastic hinge has two arms, each with an allometric cutoff, and the dynamics of hinge motion can depend critically on the synchronization of the motion of the two hinge arms, which often occurs through contacts of these cutoff maxima with other molecules. The examples treated here may not be universal, but they are biomedically critical, and their main features are easily recognized.

Bioinformatic scaling has emerged as a new thermodynamic and evolutionary method for treating allosteric interactions, with an interesting recent application to the twin enzymes cyclooxygenase (COX-1 and COX-2), isozymes encoded by separate genes [8]. Unlike traditional algebraic models, in which thermodynamics appears only after a Newtonian energy landscape has been constructed, bioinformatic scaling automatically incorporates both thermodynamics and evolution from the outset in the amino-acid-specific fractal scaling parameters $\psi$(aa) discovered by analyzing the Voronoi differential surface geometry of 5226 protein segments [9]. Surface differential geometry is relevant because the softer ice-like films of rigid proteins have evolved under environmental pressures towards self-organized criticality (evolutionary tipping points) [10,11], just as rocky coastlines are shaped fractally by the pressures of tidal water waves (discussed in Wiki articles on the coastline paradox). The power of fractal bioinformatic scaling is already apparent in its ability to quantify evolutionary trends from zebrafish to humans in both structure and function of lysozyme *c* [12,13], trends which are not accessible to Newtonian landscape models.

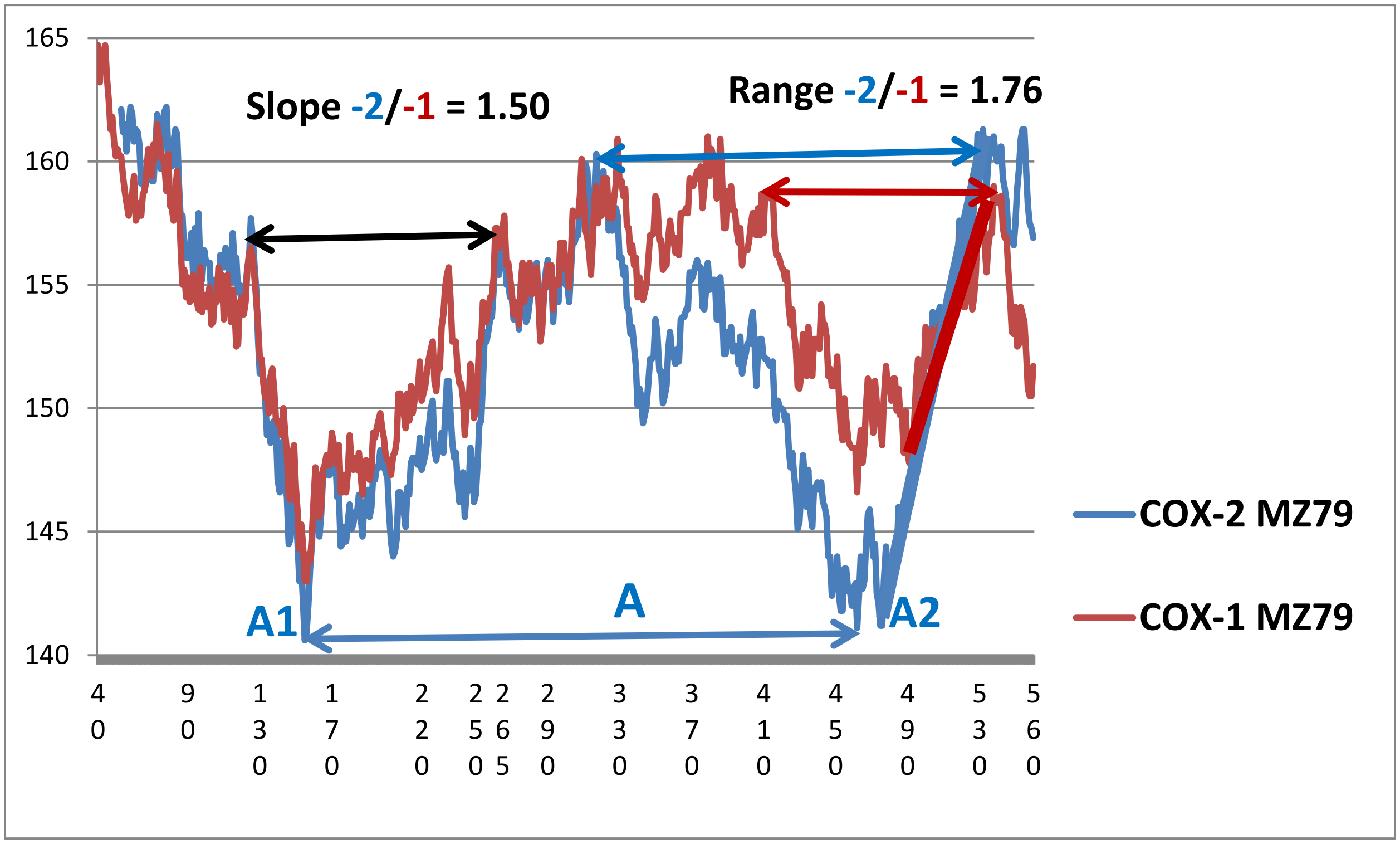

Fig. 1. Hydroprofiles for COX1/2 twins, using the fractal bioinformatic scale [5] with W = 79. Larger ψ(aa,79) ordinate values are hydrophobic, and smaller values hydrophilic. Hydroneutral on this scale is near 155. The 600 aa proteins are divided into two parts of nearly equal length separated by central hydrophobic peaks. Each N- and C- part contains a deep hydrophilic hinge, labeled A1 and A2 respectively. There are several striking differences between COX-1 and COX-2, the most obvious being the level A1 and A2 hinges of COX-2, which appear to account for its activity in restriction of platelet formation, adhesion to vessel walls, and vessel occlusion [4]. There are also large mutational effects associated with R108Q, which are thought to be allosteric, and are marked by the black double arrow centered near 200. The catalytic pocket and its entrance lie between 500 and 535, far from the A1 hinge near 160. Also note that level extrema can occur for both hydrophobic and hydrophilic extrema. As comparison with Fig. 3 shows, the A1 and A2 (N and C terminal) hinges can enable pincer motions involved in enzyme action.

Before turning to the main subject of this paper, mutations in isocitrate dehydrogenase 1 and 2 isozymes (IDH1/2) in the tricarboxylic acid cycle important for cancer cell metabolism [12], we review the central features that distinguish the COX1/2 isozymes [8]. One constructs hydroprofiles ψ(aa,W) by averaging ψ(aa) over a sliding window of width W. COX1/2 are divided hydropathically into N and C domains, each of which contains a deep hydrophilic hinge, as reproduced here from [8] for the reader's convenience (see Fig. 1). These hinges produce a large variance or roughness ℜ of the hydropathic profiles ψ(aa,W). To resolve COX1/2 differences best, W = 79 was chosen because it maximizes ℜ(COX2)/ℜ(COX1). With W = 79, Fig. 1 shows level extrema features, associated with the level set principle which has yielded many useful results for the differential geometry of other Voronoi networks [13].

**Results**

IDH1/2 are obligatory homodimeric isozymes that share 70% chain sequence identity. The dimeric structures consist of dimerized large and small domains connected by Clasps. Overall the secondary structure is ~ 40% α helical and 20% β strand, with the flexible Clasp containing only β strands. A simplified description of IDH motion and functionality is given in Fig. 3 of [14], which is accompanied by a detailed discussion of the neomorphic motion which generates the oncometabolite 2HG. Crystal structures suggest that the small/large domain interface is opened wider by IDH1 mutants than by IDH2 mutants [14]. Here we contrast IDH1 and IDH2 isozymes in two ways that involve only their amino acid sequences.

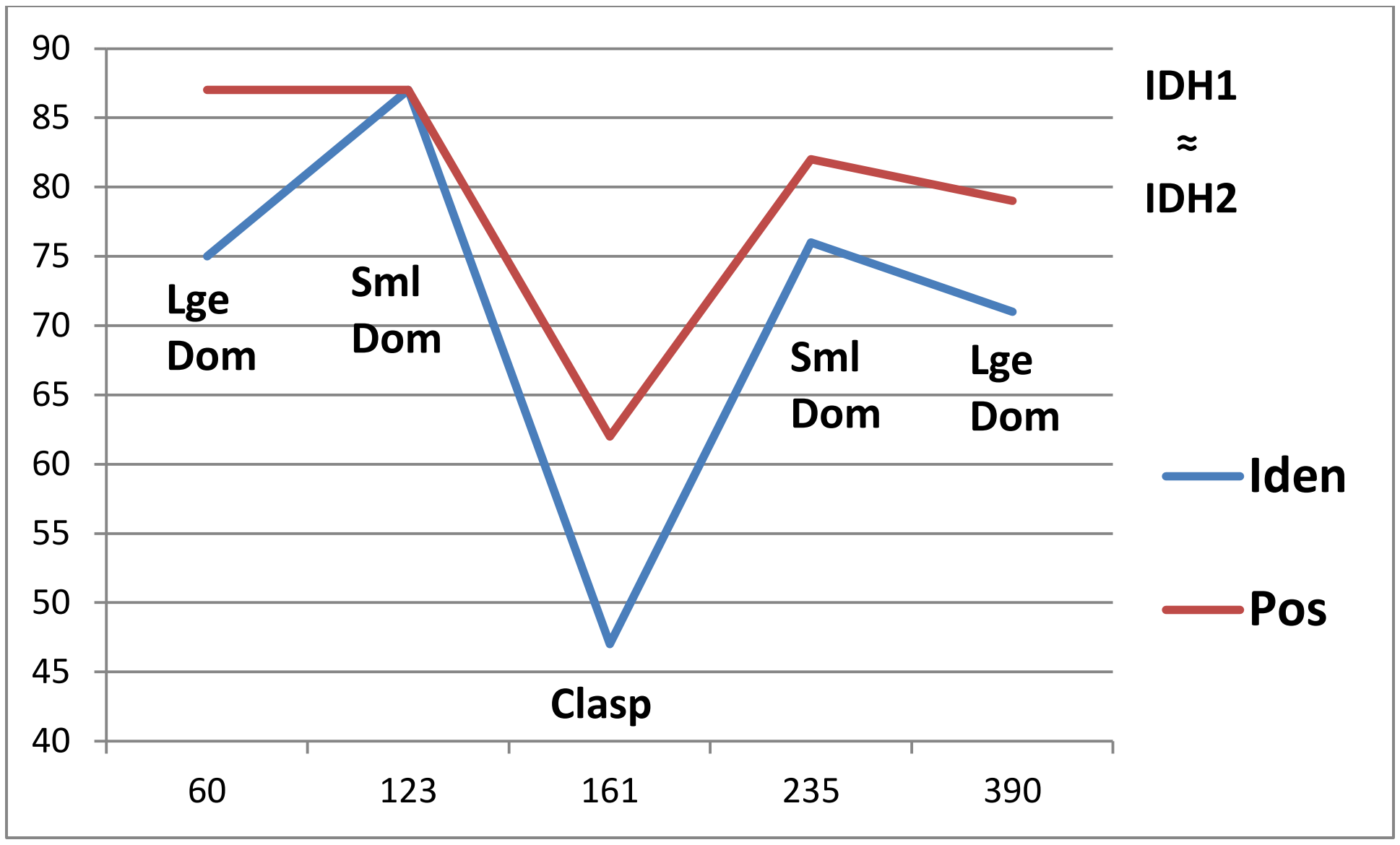

Fig. 2. BLAST sequence chain similarities 10-411 (IDH1 numbering) show weak similarity around 50% for the clasp region centered on 161, but strong similarity for the large and small domains.

First we can compare sequence identities of the three regions, as shown in Fig. 2. As expected, most of the differences between IDH1/2 chains are concentrated in the β strand Clasp region. These differences are quantified by the ψ(aa,51) profiles shown in Fig. 3. Again we see important level extrema features, so we quantify the meaning of "level". Rather than rely on a single point, we average ψ(aa,W) over nine amino acids centered on each extrema. (The number nine is chosen because this is the lower limit of the linear log-log range used to determine fractal hydropathic exponents ψ(aa) in [5].) With this definition, the IDH2 N- and C- large domain level hinge minima differ by 0.2, while the IDH1 hinge minima differ by 6.8; the range between hydrophilic and hydrophobic extrema in these units is about 25. The large domain IDH2 hinges are level to within 1%.

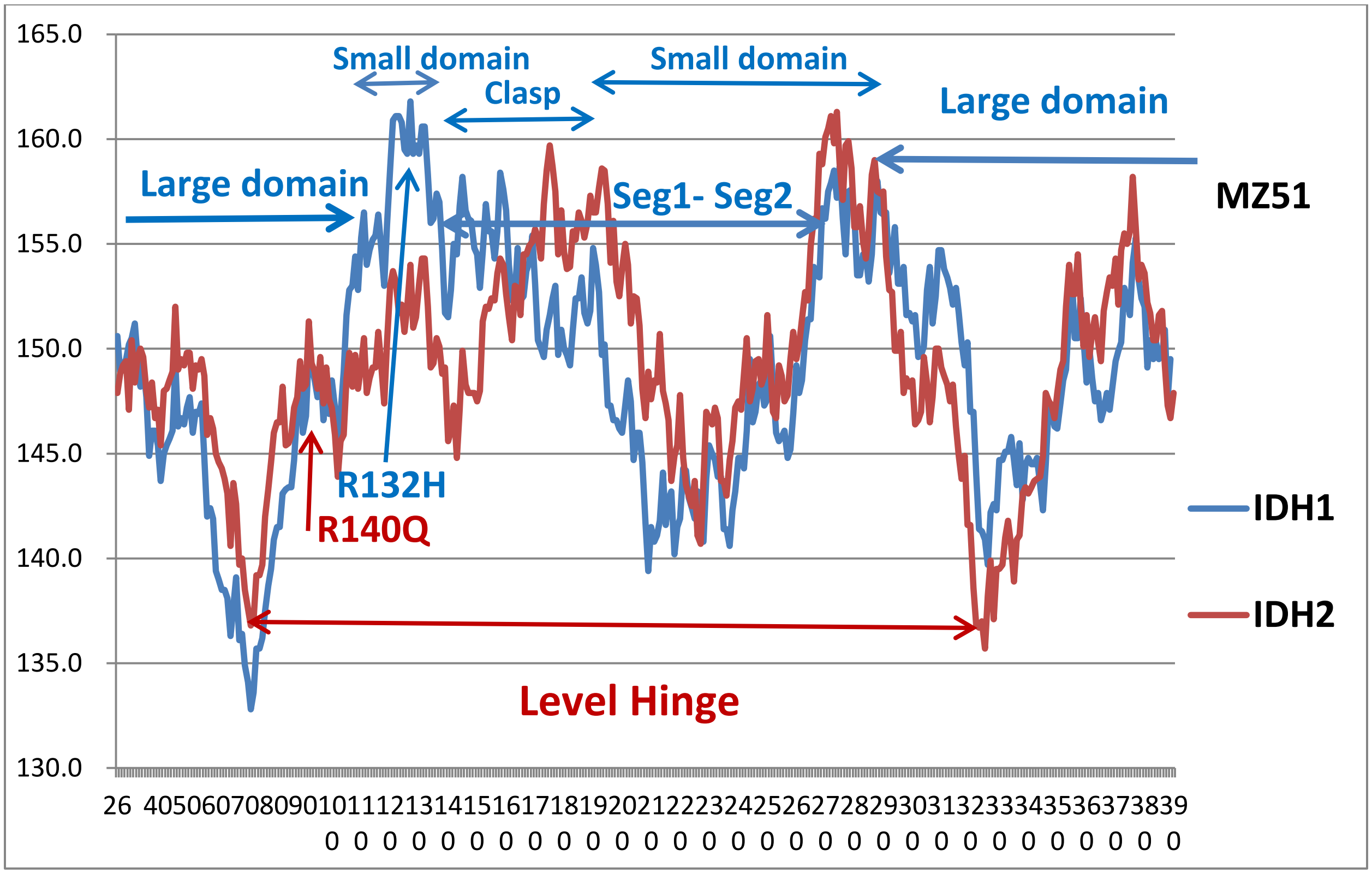

Fig. 3. The colored hydroprofiles of IDH1/2 exhibit level hydrophilic hinges between the N- and C- large domains similar to the A1-A2 hinges of COX1/2 in Fig. 1. Other IDH1/2 similarities and differences are discussed in the text, including the level Seg1/2 regions of IDH1. Sites are numbered using the IDH1 numbering, and IDH2 numbers are downshifted by 40 sites.

Another important difference between IDH1 and IDH2 is the interactions between the 132-141 (labeled Seg1) and the 271-286 (Seg2) regions of IDH1 [16]. These regions are disordered ("melted") by the mutation R132H, and have been described as part of an "off-on" switch [14]. As shown in Fig. 2A of [14], there is a salt bridge between R132 and either D275 or D279. Disruption of this salt bridge leads to formation of the Seg1/2 disordered regions. The respective averages of ψ(aa,51) for Seg1/2 are 155.6 and 156.0, so these two remote regions also are level.

So far the greatest medical interest attaches to mutated IDH2, where a very effective small molecule acute myeloid leukemia inhibitor has been discovered, Table 3 [14]. The IDH2 mutation is R140(H,Q, or C). It occurs far from the Clasp. Mutated IDH2 is treated by a small molecule that pries apart the large and small domains (Fig. 3 of [14]). At this point one can rationalize the greater biomedical success for IDH2 over IDH1 in several ways. IDH2 is simpler: it only dimerizes, while IDH1 tetramerizes. Also the R140 mutation does not induce local melting (formation of disordered Seg1,2 regions) in IDH2.

Another reason for the relative success in targeting the IDH2 R140 mutation is shown in Fig. 4; R140 is level with the N Clasp end. Replacement of hydrophilic and polar R140 with either hydroneutral or hydrophobic nonpolar amino acids (H, Q, or C) would erase this leveling by shifting ψ(aa,51) upward by 0.5 or more. This could facilitate binding to a suitable small molecule, even though R140 is only allosterically connected to the small molecule binding site. This Clasp end loosening could be a stronger and more specific "off-on" IDH2 switch than the Seg melting discussed for IDH1 [14].

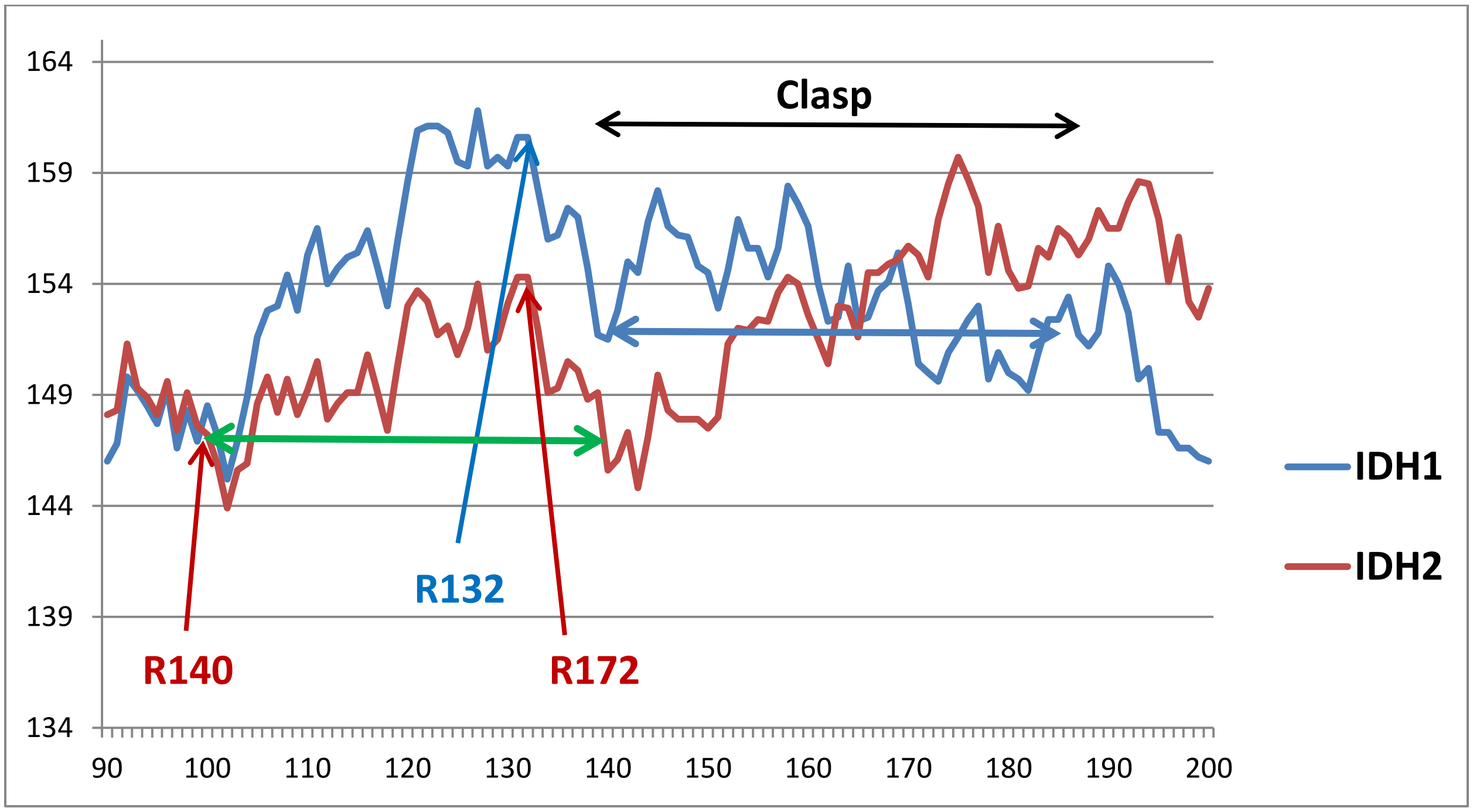

Fig. 4. Enlargement of the Clasp profile and its N terminal neighboring region, from Fig. 3. In IDH1 the Clasp ends at 140 and 186 are level, while in IDH2 the Clasp is amphiphilic, with a hydrophilic N end at 140, and a hydrophobic C end at 186. Neither the R132 site in IDH1, nor the R172 site in IDH2 are level with the Clasp ends, but R140 is level (green double arrow) with the IDH2 N Clasp hydrophilic end.

**Discussion**

As the data base on biomedical enzymes has grown, the extent of our knowledge of allosteric interactions has increased, together with the mysteries of their origin. Now by comparing Figs. 1 and 3, we can see how many such interactions can arise from modular hydropathic leveling. What is unexpected is that leveling interactions in IDH2 can be more important than salt bridges and local melting are in IDH1 [14]. This arises because of the thermodynamic and evolutionary nature of protein folding and function, as discussed at length for COX1/2 [8] and many other proteins [8,9]. One can also suppose that leveling accelerates kinetics by synchronizing allosteric conformational motions, such as N- and C–terminal bending at their hydrophilic hinges [15]. It also appears that leveling effects are most dramatic in medically significant contexts. This suggests that the final stage in optimizing protein functionality may involve leveling of the cutoffs of the arms of hinges.

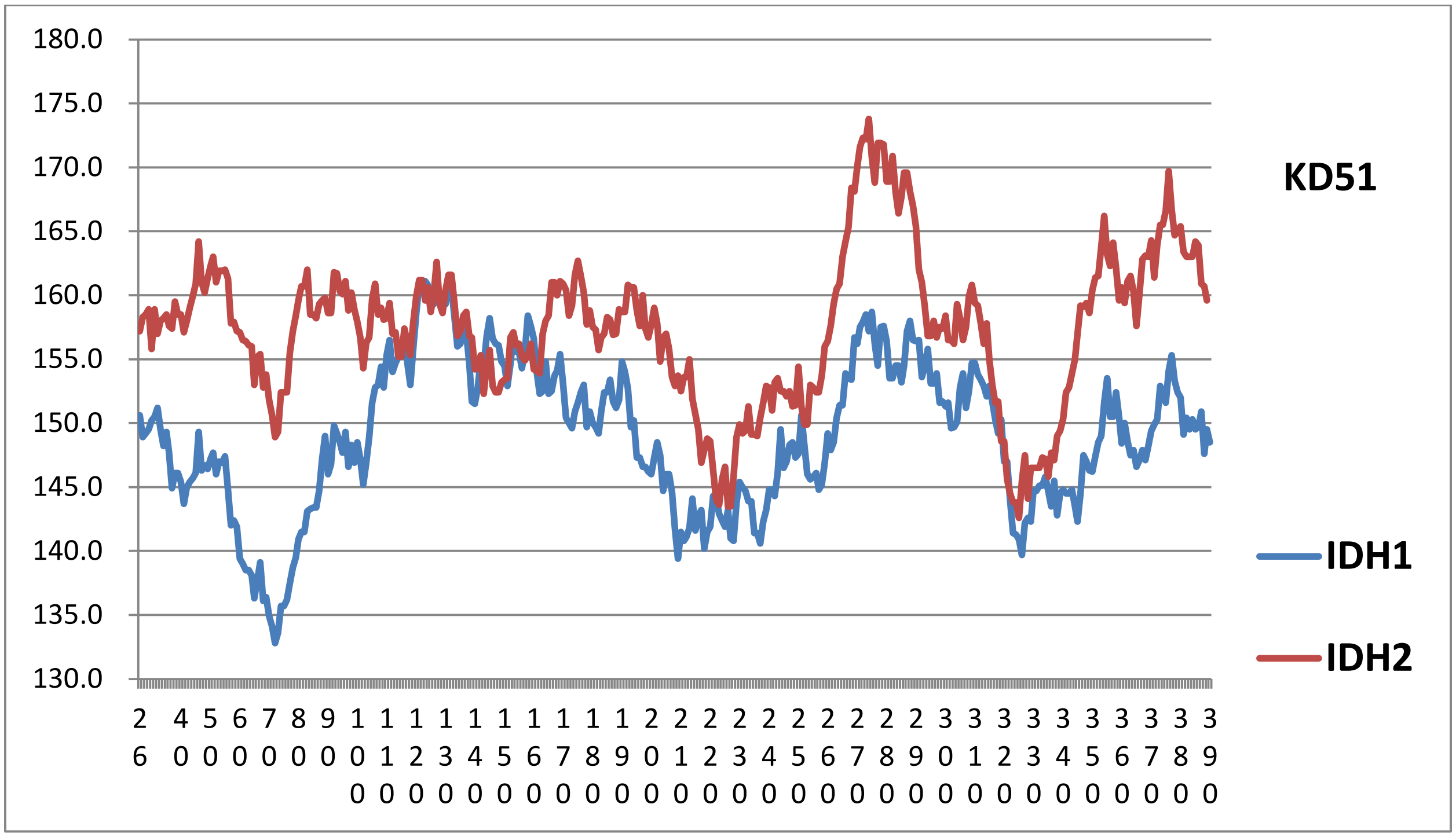

Fig. 5. This figure should be compared with Fig. 3. In Fig. 3 the two IDH1/2 isozyme profiles are quite similar, with only quantitative differences identified with level features. With the KD scale much of this similarity is lost, and of course there is little to be learned from seeking more detailed level features.

For IDH1/2 we have chosen W = 51. This choice of one scaling parameter levels many interactions, not just one, and so is overdetermined. One can illustrate the importance of overdetermined internal consistency by comparing the present profiles, based on MZ fractals [9] that describe thermodynamic conformational changes, with hydropathic profiles obtained using the coarser KD scale [16], which corresponds to complete (water-air) unfolding, and has first-order thermodynamic character. Fig. 5 shows IDH1/2 profiles with the KD scale. Although the two scales are correlated with R = 0.85, they yield qualitatively different results. Notably the close isozyme IDH1/2 similarities seen in Fig. 3 with the MZ scale are lost in Fig. 5 with the KD scale. Indeed there are more leveling similarities between COX (Fig. 1) and IDH (Fig. 3) with the MZ scale, than between IDH (Fig. 3) with the MZ scale, and IDH (Fig. 5) with the KD scale. The two enzymes both undergo similar N- and C- terminal conformational changes, although they have different folds [17,18]. This is consistent with the structural universality and functional specificity of the MZ surface fractals, derived from 5226 protein segmental geometries, which transcend fold differences [9].

The Clasp (137-185) connecting the large and small domains is a special feature of IDH proteins, and it is about 48 aa long, which is close to the value of W chosen. Furthermore, the allosteric interaction of the IDH2 R140Q mutation occurs specifically with the N end of the Clasp. This augurs well for the small molecule oncolytic drug, now in clinical trials [14], which could be specific to mutated IDH2, with acceptably small side effects due to binding to other proteins. There are intriguing similarities between the IDH Clasp and the recently discovered nicastrin lid [18], which could be discussed elsewhere in the context of amyloidosis [19,20]. The general picture of protein conformational dynamics here is similar to a dielectric model of hydration dynamics [21-23], here quantified to identify the effects of single aa mutations.

**Conclusions**

Bioinformatic scaling is based on a successful exponential compression of the sequence-structural features of the PDB into 20 parameters, made possible by self-organized criticality. The discovery of fractal hydropathic scaling parameters [9] by itself had far-reaching implications for protein structure-function relations. To develop these implications requires many specific examples, for instance, lysozyme *c* evolution of function [11,24], the panoramic evolution of thousands of flu virus strains under migration and vaccination pressures [25], classic anti-inflammatory drugs [8], amyloid fragmentation [19,20], and so on. It has also involved methods adopted from set theory, differential geometry and topology [13] to describe evolutionary flow beyond Leventhal's paradox. All of these very modern ideas seem "impossibly unlikely" from a classical Newtonian all-atom viewpoint which relies on number crunching and F = ma. Modern methods enable reliable quantification of fuzzy ideas (like network "fitness" [26]) which involve only short-range connectivity [27,28], and no long-range (allosteric) forces. The examples of allosteric resonances discussed here and in other bioinformatic scaling applications [8,11,19,25] yield "magical" results [29], which are inaccessible to classical Newtonian analysis.

**Methods**

The MZ and KD scaling parameters are tabulated in [12]. The calculations were made using an EXCEL macro written by Niels Voohoeve and refined by Douglass C. Allan.

## References

1. Kendrew JC, Bodo G, Dintzis HM, Parrish RG, Wyckoff H, Phillips DC (1958). "A Three-Dimensional Model of the Myoglobin Molecule Obtained by X-Ray Analysis". *Nature* **181** (4610): 662–666.
2. Frauenfelder H, McMahon BH, Austin RH, et al. (2001) The role of structure, energy landscape, dynamics, and allostery in the enzymatic function of myoglobin. Proc. Nat. Acad. Sci. (USA) **98**, 2370-2374.
3. Miyashita O, Onuchic JN, Wolynes PG (2003) Nonlinear elasticity, proteinquakes, and the energy landscapes of functional transitions in proteins. Proc. Nat. Acad. Sci. (USA) **100**, 12570-12.
4. Tripathi S, Portman JJ (2011) Conformational flexibility and the mechanisms of allosteric transitions in topologically similar proteins. J. Chem. Phys. **135**, 075104.
5. Uyar A, Kantarci-Carsibasi N, Haliloglu T, et al. (2014) Features of Large Hinge-Bending Conformational Transitions. Prediction of Closed Structure from Open State. Biophys. J. **106**, 2656-2666.

6. Mahajan S, Sanejouand Yves-Henri (2015) On the relationship between low-frequency normal modes and the large-scale conformational changes of proteins. Arch. Biochem. Biophys. **567**, 59-65.
7. Nguyen CN, Cruz A, Gilson MK.; et al. (2014) Thermodynamics of Water in an Enzyme Active Site: Grid-Based Hydration Analysis of Coagulation Factor Xa. J. Chem Theory Comp.**10**, 2769-2780.
8. Phillips JC (2014) Fractals and self-organized criticality in anti-inflammatory drugs. Physica A **415**, 538-543
9. Moret MA, Zebende GF (2007) Amino acid hydrophobicity and accessible surface area. Phys. Rev. E **75**, 011920.
10. Bak P, Tang C, Wiesenfeld K, Self-organized criticality – an explanation of 1/f noise. Phys. Rev. Lett. **59**, 381 (1987).
11. Phillips JC (2009) Scaling and self-organized criticality in proteins: Lysozyme *c*. Phys. Rev. E **80**, 051916.
12. LaBarre B, Hurov J, Cianchetta G, et al. Action at a Distance: Allostery and the Development of Drugs to Target Cancer Cell Metabolism (2014) Chem. & Biol. **21**, 1143-1161.
13. Saye RI, Sethian JA (2011) The Voronoi implicit interface method for computing multiphase physics. Proc. Nat. Acad. Sci. (USA) **49**, 19498-19503.
14. Yang B, Zhong C, Peng Y, et al. (2010) Molecular mechanisms of "off-on switch" of activities of human IDH1 by tumor-associated mutation R132H. Cell Res. **20**, 1188-1200.
15. Cardamone L, Laio A, Torre V, Shahapure R, DeSimone A (2011) Cytoskeletal actin networks in motile cells are critically self-organized systems synchronized by mechanical interactions. Proc. Nat. Acad. Sci. (USA) **108**, 13978-13983.
16. Kyte J, Doolittle RF A simple method for displaying the hydropathic character of a protein**.** J. Mol. Biol. **157,** 105-132 (1982).
17. Nashine VC, Hammes-Schiffer S, Benkovic SJ Coupled motions in enzyme catalysis. Curr. Opin. Chem. Biol. 14, 644-651 (2010).

18. Xie T, Yan C, Zhou R, et al. Crystal structure of the gamma-secretase component nicastrin. Proc. Nat. Acad. Sci. (USA) **111**, 13349-13354 (2014).

19. Phillips JC Thermodynamic Description of Beta Amyloid Formation. arXiv 1310.2528 (2013).

20. Phillips JC Fractal Scaling of Cortical Matter, Amyloid Fragmentation and Plaque Formation across Rodents and Primates. arXiv 1410.1419 (2014).

21. Frauenfelder H, Chen G, Berendzen J et al. A unified model of protein dynamics. Proc. Nat. Acad. Sci. (USA) **106**, 5129-5134 (2009).

22. Jansson H, Swenson J The protein glass transition as measured by dielectric spectroscopy and differential scanning calorimetry Biochim. Biophys. Acta-Prot. Prote. **1804**, 20-26 (2010).

23. Frauenfelder H Ask not what physics can do for biology - ask what biology can do for physics. Phys. Biol. **11**, 053004 (2014).

24. Phillips JC Fractals and self-organized criticality in proteins. Phys A **415**, 440-448 (2014).

25. Phillips JC (2014) Punctuated evolution of influenza virus neuraminidase (A/H1N1) under opposing migration and vaccination pressures. BioMed Research International **2014**, 907381.

26. Luksza, M; Laessig, M (2014) A predictive fitness model for influenza. Nature **507**, 57–61.

27. Caldarelli, G; Capocci, A; De Los Rios, P; et al. (2002) Scale-free networks from varying vertex intrinsic fitness. Phys. Rev. Lett **89**, 258702.

28. Boolchand P, Mauro JC, Phillips JC (2013) A note on compacted networks. Phys. Today **66**, (4) 10-11.

29. Phillips JC (2012) Hydropathic Self-Organized Criticality: A Magic Wand for Protein Physics. Protein & Peptide Letters **19**, 1089-1093.